\documentclass[nofootinbib,aps,pre,twocolumn,superscriptaddress,showkeys,showpacs,longbibliography]{revtex4-1}
\usepackage{graphicx}
\usepackage{amsmath,amssymb}

\begin{document}

\title{Intracavity Optical Trapping}

\author{Fatemeh Kalantarifard}
\affiliation{Department of Physics, Bilkent University, Ankara 06800, Turkey.}

\author{Parviz Elahi}
\affiliation{Department of Physics, Bilkent University, Ankara 06800, Turkey.}

\author{Ghaith Makey}
\affiliation{Department of Physics, Bilkent University, Ankara 06800, Turkey.}

\author{Onofrio M. Marag\`o}
\affiliation{CNR-IPCF, Istituto per i Processi Chimico-Fisici, 98158 Messina, Italy.}

\author{F. \"{O}mer Ilday}
\affiliation{Department of Physics, Bilkent University, Ankara 06800, Turkey.}
\affiliation{Department of Electrical and Electronics Engineering, Bilkent University, Ankara 06800, Turkey.}
\affiliation{UNAM -- National Nanotechnology Research Center and Institude of Material Science and Nanotechnology, Bilkent University, Ankara 06800, Turkey.}

\author{Giovanni Volpe}
\affiliation{Department of Physics, Bilkent University, Ankara 06800, Turkey.}
\affiliation{UNAM -- National Nanotechnology Research Center and Institude of Material Science and Nanotechnology, Bilkent University, Ankara 06800, Turkey.}
\affiliation{Department of Physics, University of Gothenburg, 41296 Gothenburg, Sweden.}

\date{\today}

\begin{abstract}
Standard optical tweezers rely on optical forces that arise when a focused laser beam interacts with a microscopic particle: scattering forces, which push the particle along the beam direction, and gradient forces, which attract it towards the high-intensity focal spot.
Importantly, the incoming laser beam is not affected by the particle position because the particle is \emph{outside} the laser cavity. Here, we demonstrate that \emph{intracavity nonlinear feedback forces} emerge when the particle is placed \emph{inside} the optical cavity, resulting in orders-of-magnitude higher confinement along the three axes per unit laser intensity on the sample. We present a toy model that intuitively explains how the microparticle position and the laser power become nonlinearly coupled: The loss of the laser cavity depends on the particle position due to scattering, so the laser intensity grows whenever the particle tries to escape. This scheme allows trapping at very low numerical apertures and reduces the laser intensity to which the particle is exposed by two orders of magnitude compared to a standard 3D optical tweezers. We experimentally realize this concept by optically trapping microscopic polystyrene and silica particles inside the ring cavity of a fiber laser. These results are highly relevant for many applications requiring manipulation of samples that are subject to photodamage, such as in biological systems and  nanosciences.
\end{abstract}

\maketitle

Optical tweezers are a powerful technique to manipulate microscopic particles \cite{ashkin1986observation,ashkin2000history,jones2015optical} and have found applications in various research fields, from biology \cite{fazal2011optical} and spectroscopy \cite{petrov2007raman} to statistical physics \cite{martinez2017colloidal} and nanoscience\cite{marago2013optical}. Standard optical tweezers consist of a single, typically Gaussian, beam focused by a microscope objective with a high numerical aperture (NA) \cite{jones2015optical,pesce2015step}. A microscopic particle whose refractive index is higher than that of the embedding medium can be trapped near the focal spot because of the emergence of scattering and gradient optical forces \cite{jones2015optical}.
The scattering forces are due to the radiation pressure of the light beam and act in the direction of propagation of the beam. The gradient forces push the particle towards the high-intensity focal spot. To provide gradient forces strong enough to stably trap a particle, typically water- or oil-immersion objectives with ${\rm NA}>1.20$ are used \cite{jones2015optical,pesce2015step}. Importantly, in standard optical tweezers, the laser emission is independent of the position of the particle, which is outside the laser cavity; this corresponds to an open-loop control.

Here, we demonstrate that we can dramatically enhance optical tweezers action by taking advantage of intracavity nonlinear feedback forces emerging when the microparticle is placed within the laser cavity, effectively implementing a closed-loop control. 
Surprisingly, this possibility has not been exploited until now, despite widespread interest in optical trapping and optical manipulation \cite{jones2015optical}. 
Feedback mechanisms are pervasive in science and technology, in particular linear feedback used in haptic optical tweezers \cite{pacoret2009touching}, cavity optomechanics \cite{aspelmeyer2014cavity}, laser cooling of single atoms \cite{koch2010feedback}, and recently in near-field trapping in liquid medium \cite{juan2009self,descharmes2013observation} and laser cooling in vacuum \cite{gieseler2012subkelvin}. There are also countless demonstrations of nonlinear interactions of lasers with materials leading to fascinating results \cite{menzel2013photonics}. However, the deliberate arrangement where a laser modifies a material's optical properties or position so that the material modifies the laser beam in return, constituting a nonlinear feedback mechanism, has been largely overlooked. We recently employed such arrangements to create laser-induced spatial nanostructures on various material surfaces with unprecedented uniformity \cite{oktem2013nonlinear}, to produce 3D structures deep inside silicon \cite{tokel2017nonlinear}, to demonstrate a new mechanism of highly-efficient laser material ablation \cite{kerse2016ablation}, and to obtain complex behaviour from dynamic self-assembly of colloidal nanoparticles \cite{ilday2017DSA}. We now describe an intracavity optical tweezers based on nonlinear feedback forces using  a ring-cavity fiber laser and a very low numerical aperture lens (${\rm NA} = 0.12$), which does not even permit trapping with a standard optical tweezers because of the overwhelming strength of the scattering forces that push the particle along the beam direction. 
We achieve intracavity optical trapping inside an active laser cavity where the laser mode is directly influenced by the position of the particle.
We show that this optical tweezers can stably hold microscopic objects ($3$ to $7\,{\rm \mu m}$ polystyrene and silica particles) thanks to a power self-regulation due to optomechanical coupling, obtaining a two-order-of-magnitude reduction of the average light intensity at the sample and the associated potential photodamage when compared to a standard optical tweezers that achieve the same degree of confinement. 

\subsection*{Working principle of intracavity optical trapping} 

\begin{figure}
\centerline{\includegraphics[width=.5\textwidth]{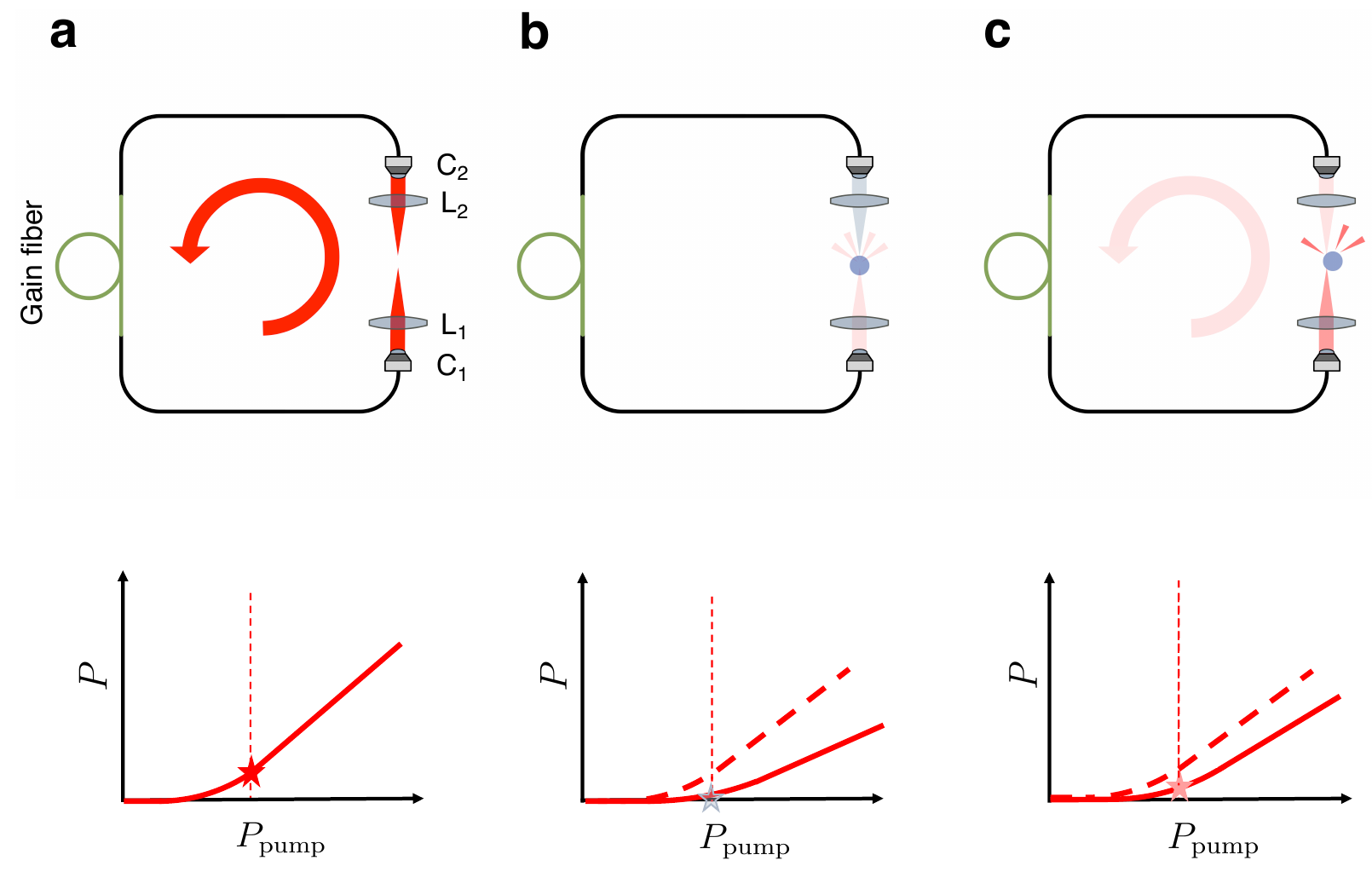}}
\caption{\textbf{Intracavity optical trapping.}
The trapping optics (collimators ${\rm C_1}$ and ${\rm C_2}$, lenses ${\rm L_1}$ and ${\rm L_2}$) are placed within the cavity of a ring fiber laser so that the position of the particle can influence the cavity loss.
(\textbf{a}) When the particle is not in the trap region, the optical loss of the cavity is low, the intracavity laser power $P$ is high, and consequently the particle is attracted towards the center of the trap. The laser power scaling curve (solid line) shows that the pump power $P_{\rm pump}$ (vertical dashed line) is above the lasing threshold.
(\textbf{b}) When the particle is at the center of the trap region, cavity losses due to scattering of light out of the cavity by the particle are maximum. The power scaling curve is right-shifted and the laser is below or barely above threshold for the same $P_{\rm pump}$. The particle is not strongly trapped. 
(\textbf{c}) When thermal fluctuations displace the particle away from the trap region, the optical loss of the cavity decreases, $P$ increases, and the particle is pulled back towards the center of the trap.}
\label{Fig1}
\end{figure}

In our scheme, the particle is trapped inside the cavity of a fiber laser, which leads to a feedback between the position of the particle and the laser emission. 
When no particle is trapped (Fig.~\ref{Fig1}a), the optical loss of the cavity is low, and the laser operates above threshold, creating a strong optical potential. When a particle is trapped (Fig.~\ref{Fig1}b), the particle scatters light out of the cavity, increasing the cavity loss and reducing the quality factor of the cavity; this leads to an increase of the lasing threshold so that the laser turns off. 
When the particle tries to escape from the trap due to thermal fluctuations (Fig.~\ref{Fig1}c), it scatters less light so that the lasing threshold decreases, the laser power increases and the particle is pulled back. 
This opto-mechanical coupling between the trapped particle and the laser cavity leads to a self-regulation of the laser power, which increases whenever the particle is about to escape, and therefore permits us to stably hold micro-objects with a very low numerical aperture lens and at a low average intensity. 

\subsection*{Toy model}

We first describe a simple toy model for our intracavity trapping scheme to clarify how the nonlinear feedback forces emerge as a result of the interplay between the particle's motion and the laser's dynamics. Like any toy model, it is designed to be simplistic, while explaining clearly and concisely the nonlinear feedback mechanism underlying the trap. It also quantifies how and to what extent this scheme reduces the average laser power to which a trapped particle is exposed. For algebraic simplicity, we discuss motion in one dimension; generalization to three dimensions is straightforward, and the results are qualitatively the same. Here, we provide an overview, while we refer to the Methods for the details. 

The motion of a Brownian particle suspended in a liquid medium and held in an attractive harmonic potential is described by the overdamped Langevin equation\cite{jones2015optical}
\begin{equation}\label{eq:langevin}
\dot{r}(t) = -\frac{k}{\gamma}r(t) + \sqrt{2D}~\xi(t),
\end{equation}
where $r(t)$ is the particle displacement from its equilibrium position, $k$ is the trap stiffness, $\gamma$ is the particle friction coefficient, $D$ is the particle diffusion coefficient, and $\xi(t)$ is a stochastic term corresponding to white noise with zero mean and unit power. The variance of the particle position in the trap is given by
\begin{equation}\label{eq:srOT}
\sigma_{\rm r, {\rm OT}}^2 = {k_{\rm B}T \over k},
\end{equation}
where $k_{\rm B}$ is the Boltzmann's constant and $T$ is the ambient temperature. The trap stiffness is proportional to the laser power,
\begin{equation}\label{eq:k}
k =\kappa_{\rm P} P, 
\end{equation}
where $\kappa_{\rm P}$ is a proportionality constant determined by the geometrical and optical properties of the setup and the sample, but independent of the laser power. For standard optical tweezers, $P$ and $k$ are independent of $r$.

\begin{figure}
\centerline{\includegraphics[width=.5\textwidth]{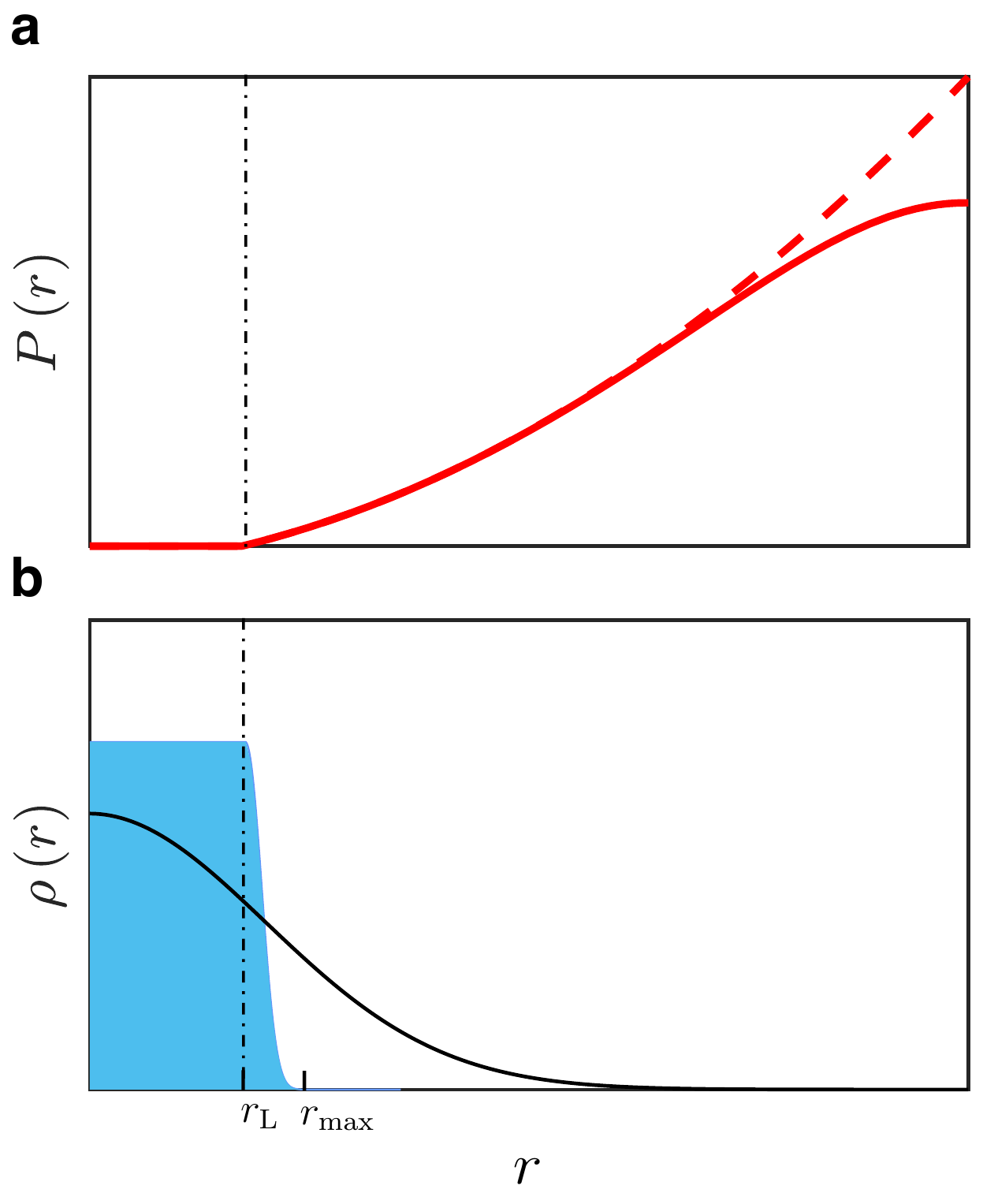}}
\caption{\textbf{Toy model of the dependence of the laser power on the particle position.}
(\textbf{a}) Laser power $P(r)$ as a function of particle position $r$ (Eq.~\ref{eq:PL}) employed in the toy model (dashed line) and the the actual laser power including saturation (solid line).
(\textbf{b}) Corresponding probability density of the particle position (Eq.~\ref{eq:rho}, with parameters $P_0 = 3\,{\rm mW}$, $r_{\rm L}=0.5\, {\rm \mu m}$, $\kappa_{\rm P} = 0.1\,{\rm pN \mu m^{-1} mW^{-1}}$, and $T=300\,{\rm K}$). The solid line represents the the probability density of the particle position obtained with a standard optical tweezers employing the same average power.}
\label{Fig2}
\end{figure}

In the intracavity optical trapping scheme, the crucial difference is that the optically trapped particle is part of the laser cavity and the particle's position modulates the loss of the cavity. We describe this coupling using a well-known model for laser dynamics by H. Haken\cite{haken2013synergetics}. We choose the laser parameters such that the net gain of the laser is negative for particle displacements smaller than a finite amount, i.e., for $r < r_{\rm L}$, so that the laser remains off. Once the particle reaches beyond $r_{\rm L}$, the laser power turns on, increasingly quadratically with $r$. Importantly, the time scale for the displacement of the particle (milliseconds) is much greater than the response time of the laser (nanoseconds), so that we can consider the laser to be always at its steady state for what concerns its effect on the particle motion. Therefore, the stationary value of the laser power is
\begin{equation}\label{eq:PL}
P(r) 
= 
\left\{
\begin{array}{ll}
0 
&
\quad r \leq r_{\rm L} 
\\
P_0
\left( \frac{r^{2}}{r_{\rm L}^{2}}- 1 \right)  
& 
\quad r >  r_{\rm L}
\end{array}
\right.\
\end{equation}
where $P_0$ is a proportionality constant. $P(r)$ is plotted by the dashed red line in Fig~\ref{Fig2}a for typical values of the parameters; the actual laser power saturates (solid line in Fig~\ref{Fig2}a), but this occurs in a region where the particle probability density is negligible and therefore is not included in the toy model.

Using Eq.~(\ref{eq:k}), the restoring force is $F(r) = - k r = -\kappa_{\rm P} P(r) r$, which is now non-harmonic because the trap stiffness depends on $r$ in the intracavity optical trapping scheme -- this positional dependency constitutes the nonlinear feedback force. 
Integrating the force, we obtain the corresponding trap potential $U(r) = -\int_0^r F(x) {\rm d}x$, and using the Boltzmann factor, the probability density of the particle position becomes
\begin{equation}\label{eq:rho}
\rho(r) 
= 
\rho_0 e^{-{U(r) \over k_{\rm B}T}}
=
\left\{
\begin{array}{ll}
\rho_{0}
& 
\quad r \leq r_{\rm L} 
\\
\rho_{0} e^{-ar^4+br^2 - {1\over2}br_{\rm L}^2}
& 
\quad r >  r_{\rm L}
\end{array}
\right.\
\end{equation}
where $a=\frac{P_0 \kappa_{\rm P}}{4 r_{\rm L}^2k_{\rm B}T}$, $b=\frac{P_0 \kappa_{\rm P}}{2 k_{\rm B}T}$, and $\rho_0$ is the normalization factor.
The blue histogram in Fig.~\ref{Fig2}b shows an example of this probability distribution for typical values of the parameters.

This model can be solved exactly, obtaining expressions for the variance of the particle position, $\sigma_r^2 = \int_0^\infty r^2 \rho(r) {\rm d}r$, and the average laser power to which the particle is exposed, $P_{\rm ave} = \int_0^\infty P(r) \rho(r) {\rm d}r$ (these exact solutions, which are rather complex, are explicitly provided in the Methods). For example, using the values of the parameters employed in Fig.~\ref{Fig2}, we obtain $\sigma_r^2 = 0.11\,{\rm \mu m^2}$ and $P_{\rm ave}=0.09\,{\rm mW}$.

We can now compare these results to the case of a standard optical tweezers with the same average power. Using Eqs.~(\ref{eq:srOT}) and (\ref{eq:k}), we obtain that 
$\sigma_{r, {\rm OT}}^2 =  {k_{\rm B}T \over \kappa_{\rm P} P_{\rm ave}}$; for the values of the parameters employed in Fig.~\ref{Fig2}, this corresponds to $\sigma_{r,{\rm OT}}^2 = 0.36\,{\rm \mu m^2}$, which represents more than 3 times less confinement than in the intracavity optical trapping scheme.
The corresponding probability distribution is shown by the solid black line plotted in Fig.~\ref{Fig2}b.

Further insight can be gained by analyzing some limiting cases that are amenable to simple analytical solutions. We will briefly consider a lower and an upper limit to the laser power exposure, providing the detailed derivation in the Methods.
We first obtain a lower limit to the average laser power by neglecting the $r^2$ term in the Maxwell-Boltzmann distribution (Eq.~(\ref{eq:rho})) so that the probability function drops faster than in the exact case. This yields 
\begin{equation}\label{eq:PaveL}
P_{\rm ave}^{\rm L}
\simeq 
\frac{2}{P_0 }
\left( \frac{k_{\rm B}T}{\kappa_{\rm P} r_{\rm L}^2} \right)^2,
\end{equation}
which corresponds to $P_{\rm ave}^{\rm L} = 0.02\,{\rm mW}$ for the parameters employed above.
Next, we obtain an upper limit by considering a uniform distribution extending to some $r_{\rm max}$, where $\rho(r_{\rm max}) \ll 1/\mathcal{C}$ with $\mathcal{C} \gg 1$ (the result is highly insensitive to the particular choice of $r_{\rm max}$):
\begin{equation}\label{eq:PaveU}
P_{\rm ave}^{\rm U}
\simeq
P(r_{\rm max}) 
\sqrt
{\ln{(\mathcal{C})}
\frac{ k_{\rm B}T}{4r_{\rm L}^2\kappa_{\rm P} P_0 }}~,
\end{equation}
which corresponds to $P_{\rm ave}^{\rm U} = 0.70\,{\rm mW}$ for the parameters employed above. 
These results show that the average power exposure is much reduced, if the laser can turn on sharply (i.e., large $P_0$ or large small-signal gain); this result motivates the choice of a fiber laser, because these lasers have extremely high small-signal gain factors (30-40 dB)  \cite{kelson1998strongly}.
Furthermore, this shows that, from an experimental perspective, we are most interested in the case of $r_{\rm L} \lesssim r_{\rm max}$. For $r < r_{\rm L}$, the laser is off and the particle experiences free diffusion with a uniform probability density. To take advantage of intracavity trapping, we are interested in parameter combinations for which free diffusion is dominant. In this case, the laser has to be off or operating at low power up to some large $r_{\rm L}$ and be turned on quickly for $r > r_{\rm L}$, so as to erect a steep potential barrier that confines the particle. 

\subsection*{Simulations}

\begin{figure}
\centerline{\includegraphics[width=.5\textwidth]{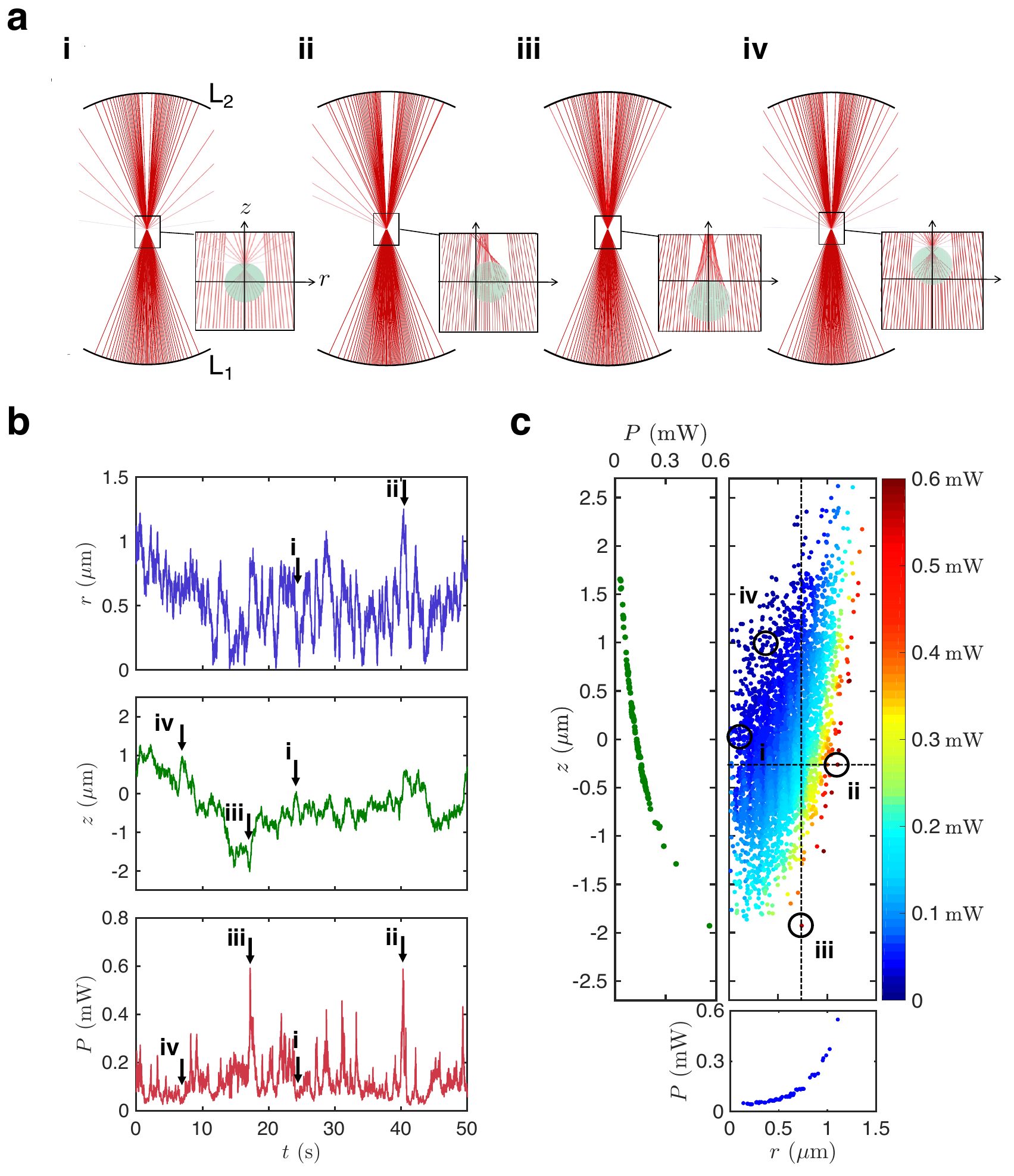}}
\caption{\textbf{Simulations.}
(\textbf{a}) Ray optics diagrams of the propagation of a focused beam through an optically trapped particle
(i) when the particle is at the center of the trap (equilibrium position in the trap that takes into account also the effective gravitational force acting on the particle),
(ii) when it is displaced in the radial direction,
(iii) when it is displaced along the axial direction downwards
and (iv) upwards.
(\textbf{b}) Radial ($r$) and axial ($z$) particle position, and corresponding laser power ($P$) obtained from the simulation of the motion of a $4.9$-${\rm \mu m}$-radius polystyrene particle trapped in the intracavity optical trap. 
(\textbf{c}) Dependence of the laser power on the radial and axial position of the particle. The points (i)-(iv) correspond to the configurations in (a) and the dashed lines correspond to the insets graphs the dependence on $z$ and $r$ on the left and bottom respectively.}
\label{Fig3}
\end{figure}

We now present a series of numerical simulations based on an extended theoretical model, including highly realistic descriptions of the laser dynamics, optical losses incurred by the particle, and the particle's Brownian motion in order to gain a quantitative understanding of the dynamics of intracavity optical trapping and to guide the experiments. The dynamics of the trapped particle are similarly governed by the interplay between the optical force $\textbf{F}_{\rm ot}(\textbf{r},P)$, the gravity minus the buoyancy $\textbf{F}_{\rm g}$, the viscous drag acting on the particle, and the thermal fluctuations.
The resulting overdamped Langevin equation is \cite{jones2015optical}:
\begin{equation}\label{eq:ot}
\dot{\textbf{r}}=\gamma^{-1} \left[ \textbf{F}_{\rm ot}(\textbf{r},P) + \textbf{F}_{\rm g} \right] + \sqrt{2\textit{D}}~\textbf{W}(t)~,
\end{equation}
where $\textbf{r}$ is the particle position, $\gamma=6\pi \eta R$ is its friction coefficient that depends on the particle radius $R$ and the medium viscosity $\eta$, $P$ is the power of the laser, $D={k_{\rm B}T}/\gamma$ is the particle diffusion that depends on the temperature $T$, $k_{\rm B}$ is the  Boltzmann constant, and $\textbf{W}(t)$ is a vector of independent white noises.  

The laser dynamics is modeled using standard power rate equations \cite{kelson1998strongly} (see Methods for details). This is a highly realistic model that includes gain saturation, which was ignored in the toy model. 
We note that the characteristic timescale for particle displacement due to Brownian motion is in the milliseconds, whereas the laser dynamics is in the nanosecond scale. Thus, we take the laser to be always at its steady state and calculate its power, given the loss corresponding to the particle position.

Since the size of the particles is significantly larger than the wavelength \cite{ashkin1992forces}, we have calculated the trapping force, as well as the loss, using the geometrical optics approach implemented in the software package Optical Tweezers in Geometrical Optics (OTGO) \cite{callegari2015computational}: The incoming laser beam is decomposed into a set of optical rays, which are then focused by the focusing lens. As the rays reach the particle, they get partially reflected and partially transmitted. The direction of the reflected and transmitted rays are different from those of the incoming rays. This change of direction entails a change of momentum and a force acts on the particle due to the action-reaction law. If the refractive index of the particle is greater than that of the medium as is usually the case, these optical forces tend to pull the sphere towards a stable equilibrium position near the focal spot. The optical forces are proportional to the laser power. As the scattered rays reach the collecting lens, they are collected and projected onto the back-focal plane input of the fiber.

Fig.~\ref{Fig3} illustrates the simulations results for a $4.9$-${\rm \mu m}$-radius polystyrene particle held in an intracavity optical trap. The particle confinement is $\sigma_r^2 = 0.18\,{\rm \mu m^2}$ and $\sigma_z^2 = 0.38\,{\rm \mu m^2}$, which are in good agreement with the results obtained for the experiments presented in the next section.
Some interesting configurations of particle and laser are shown in Fig.~\ref{Fig3}a, corresponding to the positions indicated in Fig.~\ref{Fig3}b.
When the particle is in the center of the trap, a significant part of the light is scattered out of the collector lens, as can be seen in the ray optics diagram shown in panel i of Fig.~\ref{Fig3}a, which leads to a low laser power (position i in Fig.~\ref{Fig3}b).
When the particle moves away from the trap along the radial direction (panel ii of Fig.~\ref{Fig3}a), this leads to an increase of the laser power that reaches the collector lens (position ii in Fig.~\ref{Fig3}b) and, therefore, to an increased restoring force pulling back the particle. 
Similarly, when the particle moves down along the axial direction (panel iii of Fig.~\ref{Fig3}a), there is an increase of the collected power (position iii in Fig.~\ref{Fig3}b) and an increase of the scattering force that pushes the particle up.
Finally, when the particle moves up along the axial direction (panel iv of Fig.~\ref{Fig3}a), the rays are scattered away from the collector lens (position iv in Fig.~\ref{Fig3}e) and this decreases the laser power letting the particle fall back towards the center of the trap.

To better understand the coupling between the particle position fluctuations and the laser power, we have plotted the laser power as a function of the radial and axial position of the particle in Fig.~\ref{Fig3}c.
The power increases whenever the radial position $r$ increases, confirming that, when the particle tries to escape along the radial direction, the laser increases its power and pulls back the particle towards the center of the trap. 
When the particle moves down, the laser power increases and the scattering force pushes the particle up towards the trap center, and, when the particle moves up, the laser decreases the power and the scattering force, which permits the particle to fall down towards the center of the trap.

\subsection*{Experimental results}

\begin{figure}
\centerline{\includegraphics[width=.5\textwidth]{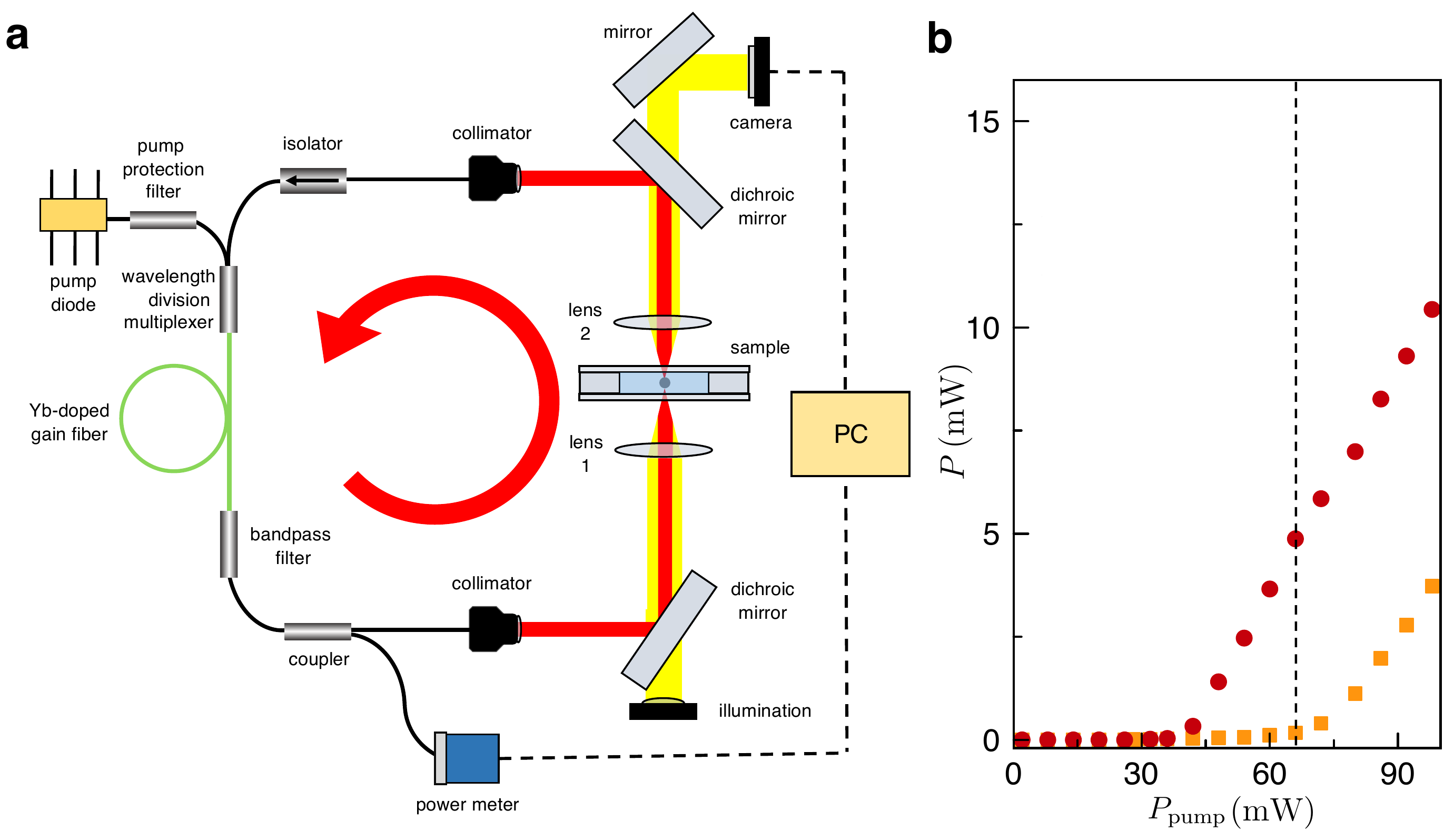}}
\caption{\textbf{Experimental setup}. 
\textbf{(a)} The setup comprises a diode-pumped Yb-doped fiber laser, the trapping optics, and the digital video microscope.
\textbf{(b)} Measured power scaling with a trapped $4.9$-${\rm \mu m}$-radius polystyrene particle (orange squares) and without the trapped particle (red circles). At a pump power of $66\,{\rm mW}$ (dashed vertical line), the laser is below threshold with the particle (orange squares), but above threshold without the particle (red circles).}
\label{Fig4}
\end{figure}

Finally, guided by the simulation results, we have built an experimental setup to prove the operational principle of intracavity optical trapping (Fig.~\ref{Fig4}a). We constructed a continuous-wave ring-cavity fiber laser emitting at $1030\,{\rm nm}$ with the trapping optics placed inside the cavity. The laser beam is directed
upward and the trap is achieved by a lens with an effective ${\rm NA}=w/f=0.12$, where $w=1.0\,{\rm mm}$ is the beam waist at the lens and $f=8.0\,{\rm mm}$ is the lens focal length. 

The cavity further comprises a single-mode Yb-doped fiber (Yb 1200-6/125, nLIGHT, Inc., core diameter of $6\,{\rm \mu m}$, cladding diameter of $125\,{\rm \mu m}$) as gain medium, pumped by a single-mode fiber-pigtailed diode-laser at $976\,{\rm nm}$ through a wavelength division multiplexer. To achieve spectral stability, we included an fiberized band-pass filter (centered at $1030\,{\rm nm}$, full-width at half wavelength of $2\,{\rm nm}$, placed after the gain fiber). After the band-pass filter, the beam is split by a coupler with 99:1 coupling ratio: the larger portion is sent to a single-mode fiber-pigtailed collimator (OZ Optics, Ltd.), and the smaller portion is sent to a photodiode power sensor (S150C, Thorlabs, Inc., $10\,{\rm pW}$ resolution) for power monitoring. The output of the collimator is reflected by a short-pass dichroic mirror (DMSP1000, Thorlabs, Inc.) and focused on the sample by an aspheric lens ($8.0\,{\rm mm}$ focal length, ${\rm NA}=0.12$). The laser light is then collected by a second identical aspheric lens, reflected by a short-pass dichroic mirror (DMSP1000, Thorlabs, Inc.), and coupled back into the fiber by a collimator. An in-line isolator is used to ensure that the light is traveling within the cavity unidirectionally. The roundtrip time of the cavity is $\simeq 20\,{\rm ns}$  and total losses are $50\%$, which implies that the cavity responds to change in losses in the scale of $40\,{\rm ns}$, which is practically instantaneous compared to the Brownian motion of the trapped particle. The red circles in Fig.~\ref{Fig4}b show the power scaling curve of the laser as a function of the pump power.

We have suspended polystyrene (radii of $4.9\,{\rm \mu m}$ and $6.2\,{\rm \mu m}$,  Microparticles, GmbH) or silica (radii of $2.8\,{\rm \mu m}$, $4.0\,{\rm \mu m}$ and $4.8\,{\rm \mu m}$,  Microparticles, GmbH) particles in water and placed a droplet of the resulting solution in a sample chamber realized between two microscope slides separated by a parafilm layer ($100$-${\rm \mu m}$ thick). We placed the sample on a 3-axis translation stage.
The orange squares in Fig.~\ref{Fig4}b show the power scaling curve of the laser as a function of the pump power when a $4.9$-${\rm \mu m}$-radius polystyrene particle placed at focal point: at the pump power of $66\,{\rm mW}$, the laser power is reduced from $5.0\,{\rm mW}$ to $0.2\,{\rm mW}$.

For imaging, we have illuminated the sample using a LED ($630\,{\rm nm}$ wavelength, $20\,{\rm nm}$ bandwidth) whose coherence length is much shorter than the thickness of the microscope slides and the separation between the two slides to avoid interference. We have recorded the motion of the particle using a CMOS camera (DCC1645C, Thorlabs, Inc., $50\,{\rm Hz}$ frame rate) and used digital video microscopy to track its trajectory in 3D (see Methods).

\begin{figure}
\centerline{\includegraphics[width=.5\textwidth]{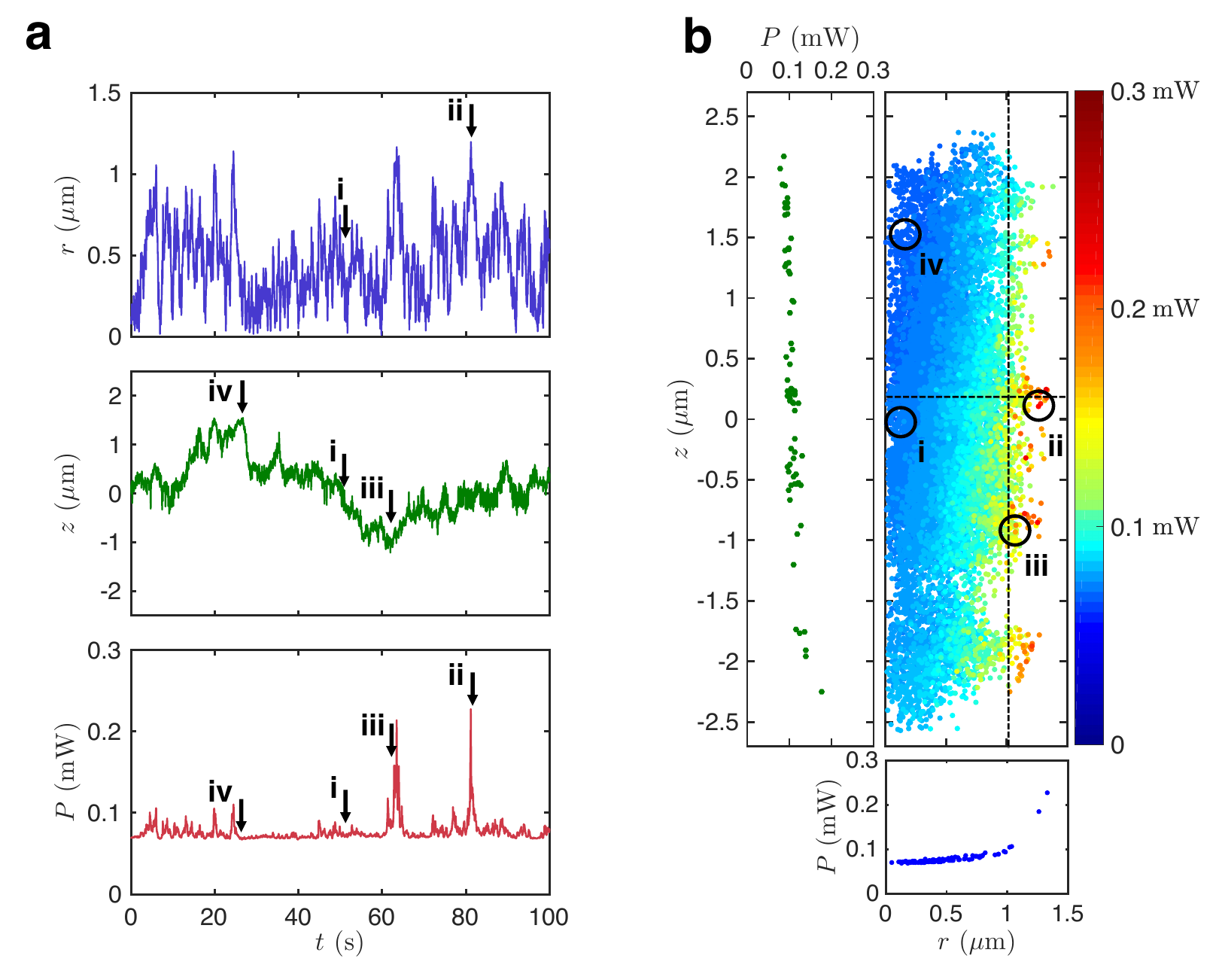}}
\caption{\textbf{Experiments.}
(\textbf{a}) Radial ($r$) and axial ($z$) particle position, and corresponding laser power ($P$) for a $4.9$-${\rm \mu m}$-radius polystyrene held in the intracavity optical trap. 
(\textbf{b}) Dependence of the laser power on the radial and axial position of the particle.
The positions indicated with i, ii, iii and iv correspond to particle position and laser power configurations similar to those illustrated in the diagrams shown in Fig.~\ref{Fig3}a. The points (i)-(iv) correspond to the configurations shown in Fig.~{Fig3}a, and the dashed lines correspond to the insets graphs the dependence on $z$ and $r$ on the left and bottom respectively.
}
\label{Fig5}
\end{figure}

In Fig.~\ref{Fig5}, we show the results we obtained by trapping a $4.9$-${\rm \mu m}$-radius polystyrene particle at a pump power of $66\,{\rm mW}$ corresponding to $5.0\,{\rm mW}$ laser output in the absence of the particle.
Fig.~\ref{Fig5}a shows the time evolution of the particle radial position $r$, its axial position $z$, and the laser power $P$.
We have measured the confinement achieved by the trap in the radial and axial directions calculating the variance of the particle position, which are $\sigma_r^2 = 0.038 \,{\rm \mu m^2}$ and $\sigma_z^2 = 0.41 \,{\rm \mu m^2}$, which are in good agreement with the results of the simulations presented in Fig.~\ref{Fig3}.
Fig.~\ref{Fig5}b shows how the laser power depends on the radial and axial particle position, which is also in agreement with the results of the simulations presented in Fig.~\ref{Fig3}c. Along the radial direction, the laser power increases when $r$ is large, providing enough restoring force to push the particle back towards the center of the trap. Along the axial direction, the laser power decreases when the particle moves upwards leading to the particle moving downwards because of sedimentation, and increases when the particle moves downwards, leading to enhanced scattering forces pushing the particle back upwards.

\subsection*{Comparison with standard optical tweezers}

\begin{figure}
\centerline{\includegraphics[width=.5\textwidth]{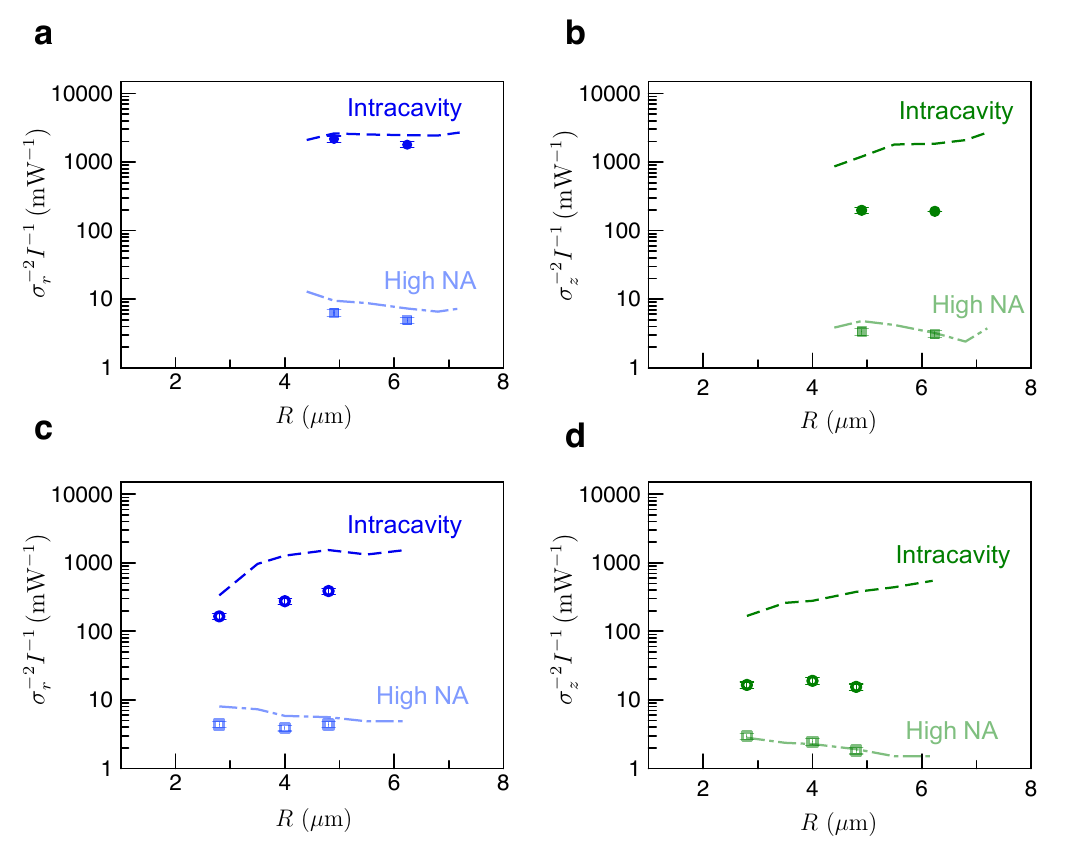}}
\caption{\textbf{Enhancement of particle confinement per unit intensity at the sample.} 
Comparison of the experimentally measured inverse radial and axial trap confinement ($\sigma_r^{-2}$ and $\sigma_z^{-2}$, respectively) per unit intensity at the sample for an intracavity optical trap (circles) and a standard high-NA optical tweezers (squares) for polystyrene (\textbf{a}-\textbf{b}) and silica (\textbf{c}-\textbf{d}) particles of various radii $R$.
The dashed lines are the corresponding results from numerical simulations.}
\label{Fig6}
\end{figure}

It is instructive to compare intracavity optical trapping to a standard optical tweezers. To this end, we have compared the inverse radial and axial confinement, $\sigma_r^{-2}$ and $\sigma_z^{-2}$, per unit intensity at the sample, $I$, measured using the intracavity optical tweezers to that obtained using standard optical tweezers with low and high numerical apertures.

For the case of low-NA standard optical tweezers (${\rm NA}=0.12$), we observe that the particle is never trapped along the axial direction because the restoring force is not strong enough. Therefore, intracavity optical trapping provides a simple, self-aligning technique to achieve trapping at low NA. 
We remark that several alternative approaches have been proposed to make optical tweezers capable of trapping particles at low NA, but all these methods require the use of multiple/structured optical beams or special sample preparation, e.g. counter-propagating optical tweezers \cite{vcivzmar2011holographic}, mirror trapping \cite{pitzek2009optical}, focused Bessel beams \cite{ayala20163d}, self-induced back action \cite{juan2009self,descharmes2013observation}.

For the case of high-NA standard optical tweezers (${\rm NA}=1.30$), we trapped a $4.9$ ${\rm \mu m}$-radius polystyrene particle with a standard optical tweezers using a high-NA microscope objective (${\rm NA}=1.30$) using a laser with a central wavelength of $976\,{\rm nm}$. We measured the particle trajectories using digital video microscopy and analyzed them to obtain the trap stiffness.
We measured $\sigma_{{\rm hNA},r}^{-2}/I_{\rm hNA} = 6.3\,{\rm mW^{-1}}$ and $\sigma_{{\rm hNA},z}^{-2}/I_{\rm hNA} = 3.7\,{\rm mW^{-1}}$, where $I_{\rm hNA} = 35\,{\rm mW\, \mu m^{-2}}$ is the intensity at the sample.
Instead, the intracavity optical trap achieves about two order of magnitude tighter confinement for the same intensity at the sample: $\sigma_r^{-2}/I =2180\,{\rm mW^{-1}}$ and $\sigma_z^{-2}/I = 200\,{\rm mW^{-1}}$, where $I = 0.012\,{\rm mW\, \mu m^{-2}}$ is the intensity at the sample.

Fig.~\ref{Fig6} shows a comparison of the experimental and simulated confinement per unit intensity at the sample for polystyrene and silica particles of various sizes. The intracavity confinement per unit intensity are consistently about two orders of magnitude higher than standard optical tweezers values.

\subsection*{Conclusions}

We have demonstrated, through a simple analytical model, simulations of a realistic model, and experiments, a novel optical trapping scheme, where the laser operation is nonlinearly coupled to the motion of the trapped particle. This coupling gives rise to nonlinear feedback forces. We implemented this scheme within the cavity of a fiber laser because its large, small signal gain is essential for achieving a high trap stiffness per unit intensity. We have trapped polystyrene and silica microparticles of different sizes. 
The main advantages of this scheme are that it can operate with very low-NA lenses and at very low average power, resulting in about two orders of magnitude reduction in exposure to laser intensity compared to standard optical tweezers. We anticipate that intracavity optical traps, enabling 3D confinement with long-working-distance lenses, low operating intensity, simple optics and low costs can prove useful in several research fields, particularly in biology where low photodamage of the sample is often crucial.

\section*{Methods}

\subsection*{Detailed description of the toy model}  

\subsubsection*{Model for laser dynamics (Eq.~(\ref{eq:PL})).}

We describe the laser dynamics using a model based on the one introduced by H. Haken\cite{haken2013synergetics}
\begin{equation}\label{eq:haken}
\dot{P} = NWP - \frac{l(r)}{\tau_{\rm R} } P~,
\end{equation}
where $N = N_{0} - \frac{2D_{0}\tau}{h\nu}P$ is the electron population in the excited state, $N_{0}$ is the  population in the excited state determined by the pump power, $W$ is the spontaneous emission rate at which excited electrons relax to the ground state, $\tau$ is the relaxation time, $\tau_{\rm R}$ is the cavity roundtrip time, $h$ is the Planck's constant, $\nu$ is the optical frequency of the laser output, and $l(r)$ is the loss of the cavity, which depends on the particle position $r$. Inserting $N$ into the equation above gives
\begin{equation}
\dot{P} = G_{\rm net}(r)P - \frac{2N_{0}\tau W}{h\nu}P^2,
\end{equation}
where $G_{\rm net}(r) = N_{0}W - l(r) / \tau_{\rm R}$ is the net gain, which depends on the particle position. While the dependence of the losses on the particle's position is, in general, complex, it is maximum at the center and drops quadratically for small displacements as $l(r) = l_0 (1- r^2/r_{\rm c}^2)$, where $r_{\rm c}$ characterizes the scaling of the losses with the particle's position, which depends on the radius of the particle, relative indices of refraction of the particle and the liquid medium, and the losses due to absorption and scattering. The laser cavity is intentionally arranged to be below threshold (i.e., $G_{\rm net}(r)<0$) for $r < r_{\rm L}$, where $r_{\rm L}$ is the characteristic particle displacement where lasing turns on. Therefore, the stationary value of the laser power is given by Eq.~(\ref{eq:PL}), i.e.
$$
P(r) 
= 
\left\{
\begin{array}{ll}
0 
& 
\quad r \leq r_{\rm L} \\
P_0 \left( \frac{r^{2}}{r_{\rm L}^{2}}- 1 \right)  
& 
\quad r >  r_{\rm L}
\end{array}
\right.
$$
where $G_0=l_0 / \tau_{\rm R} -N_{0}W $, $r_{\rm L}=r_{\rm c}\sqrt{G_0\tau_{\rm R}/l_0}$, and $P_0 = h\nu G_0 / (N_0\tau W)$, which is the the normalized slope of laser power with respect to the particle position and depends on laser parameters, such as optical loss, gain characteristics, and pump power.
A more complete model would incorporate gain saturation as well as the saturation of the decrease of losses with increasing $r$, which ultimately limits the laser power to a finite value. Nevertheless, we can use this algebraically simpler model because the probability of the particle to be displaced to large values of $r$, where saturation matters, decreases exponentially. 

\subsubsection*{Stationary probability density of the particle position (Eq.~(\ref{eq:rho})).}

The restoring force is given by
\begin{equation}
F(r) 
=
- k r
=
-\kappa_{\rm P} P(r) r
=
\left\{
\begin{array}{ll}
0 
& 
\quad r \leq r_{\rm L} \\
-P_0 \kappa_{\rm P}
\left( \frac{r^{2}}{r_{\rm L}^{2}} - 1 \right) r
& 
\quad r >  r_{\rm L}
\end{array}
\right.
\end{equation}
This is an important result that warrants several comments. 
First, it is the dependence of trap stiffness on position that constitutes a nonlinear feedback mechanism. 
Second, the particular form of this nonlinear response is such that the laser power is zero up to a certain displacement, $r_{\rm L}$, and increases quadratically afterwards (until reaching saturation, which is ignored here). We will show below that this particular form results in a much reduced average power that a trapped particle experiences, compared to having fixed laser power as in a traditional trap for the same level of confinement (i.e., variance in the trap). 
Integrating the force, the corresponding trap potential is found to be
\begin{equation}
U(r) 
=
-\int_0^r F(x) dx
=
\left\{
\begin{array}{ll}
0 
& 
\quad r \leq r_{\rm L} \\
P_0 \kappa_{\rm P} \left( \frac{r^4}{4r_{\rm L}^2} - \frac{r^2}{2} + \frac{r_{\rm L}^2}{4} \right)
& 
\quad r >  r_{\rm L}
\end{array}
\right.
\end{equation}
From the potential, we use the Boltzmann factor to determine the probability density of the particle position, obtaining Eq.~(\ref{eq:rho}).

\subsubsection*{Analytical expressions for $\sigma_r^2$ and $P_{\rm ave}$.}

We can calculate the variance of the particle position as
\begin{equation}\label{eq:sr} 
\sigma_{\rm r}
=
\int_0^{\infty} r^2 \rho(r) dr
=
\int_{0}^{r_{\rm L}}r^2\rho_{0} {\rm d}r +\int_{r_{\rm L}}^{\infty}r^2\rho_{0}e^{-ar^4+br^2-br_{\rm L}^2/2} {\rm d}r.
\end{equation}
and the average laser power value to which the particle is exposed as
\begin{equation}\label{eq:Pave}
\begin{split}
P_{\rm ave}
& =
\int_0^{\infty} P(r) \rho(r) dr \\
& =
\int_{r_{\rm L}}^{\infty}P_0
\left(
\frac{r^2}{r^2_{L}}-1
\right) 
\rho_{0}e^{-ar^4+br^2
-
br_{\rm L}^2/2} {\rm d}r.
\end{split}
\end{equation}
The solutions of integrals in Eqs.~(\ref{eq:Pave}) and (\ref{eq:sr}) can be expressed as series: If we expand $e^{br^2}$ in a Taylor series, with its convergence guaranteed by the faster decaying term, $e^{-ar^4}$, we can exactly evaluate the integrals corresponding to each term using $\int_{r_{\rm L}}^{\infty}r^{2n}e^{-ar^4} {\rm dr}=\frac{\Gamma\left(\frac{2n+1}{4},ar_{L}^4\right)}{4a^{\frac{2n+1}{4}}}$. Thus, the variance and the average power exposure are
\begin{equation}
\sigma_{\rm r}=\frac{
 \frac{r_{\rm L}^3}{3}+e^{-br_{\rm L}^2/2}\sum_{n=0}^\infty \frac{b^n\Gamma\left(\frac{2n+3}{4},ar_{L}^4\right)}{4a^{\frac{2n+3}{4}}n!}
}
{r_{\rm L}+e^{-br_{\rm L}^2/2}\sum_{n=0}^\infty 
\left(
\frac{P_0 b^n\Gamma\left(\frac{2n+1}{4},ar_{L}^4\right)}{4a^{\frac{2n+1}{4}}n!}
\right) }
\end{equation}
and
\begin{equation}
P_{\rm ave}=\frac{e^{-br_{\rm L}^2/2}
\sum_{n=0}^\infty 
{\left(
\frac{P_0 b^n\Gamma\left(\frac{2n+3}{4},ar_{L}^4\right)}{4a^{\frac{2n+3}{4}}r_{\rm L}^2n!}-\frac{P_0 b^n\Gamma\left(\frac{2n+1}{4},ar_{L}^4\right)}{4a^{\frac{2n+1}{4}}n!}
\right)}}{r_{\rm L}+e^{-br_{\rm L}^2/2}\sum_{n=0}^\infty 
\left(
\frac{P_0 b^n\Gamma\left(\frac{2n+1}{4},ar_{L}^4\right)}{4a^{\frac{2n+1}{4}}n!}
\right)}~,
\end{equation}  
where $\Gamma\left(s,x\right)$ is the upper incomplete gamma function  defined as $\Gamma\left(s,x\right)=\int_x^\infty u^{s-1}e^{-u}{\rm d}u$ \cite{abramowitz1964handbook}.

\subsubsection*{Exposure to laser power: lower limit (Eq.~(\ref{eq:PaveL})).}

We can calculate a lower limit to the average laser power to which the particle is exposed by neglecting the $r^2$ term in the expression for $\rho(r)$ (Eq.~(\ref{eq:rho})). Compared to the exact form, the probability function drops faster without the $r^2$ term, ensuring that such a calculation constitutes a lower limit. We therefore obtain
\begin{equation}      
\sigma_{{\rm L}, r}^2
\simeq 
\rho_{0}^{'}
\int_{0}^{r_{\rm L}}r^2{\rm d}r 
+
\rho_{0}^{'} \int_{r_{\rm L}}^{\infty}r^2 e^{-ar^4+br_{\rm L}^2/2} {\rm d}r
\end{equation}
and
\begin{equation}      
P_{\rm ave}^{\rm L}
\simeq 
\rho_{0}^{'}
\int_{r_{\rm L}}^{\infty}P_0 
\left(
\frac{r^2}{r^2_{L}}
-1
\right) e^{-ar^4+br_{\rm L}^2/2} {\rm d}r~,
\end{equation}
where $\rho_0^{'}
=
\left[{\int_{0}^{r_{\rm L}}{\rm d}r  + \int_{r_{\rm L}}^{\infty}e^{-ar^4+br_{\rm L}^2/2}}{\rm d}r \right]^{-1}
=
\left[ r_{L} + \frac{e^{br_{\rm L}^2/2}}{4a^{\frac{1}{4}}} \Gamma\left(\frac{1}{4},ar_{\rm L}^4\right] \right)^{-1}$. These integrals can be evaluated using the upper incomplete gamma functions, obtaining
\begin{equation}      
\sigma_{{\rm L}, \rm r}^2
\simeq
\rho_{0}^{'}
\left[
\frac{r_{\rm L}^3}{3}
+\frac{e^{br_{\rm L}^2/2}}{4a^{\frac{3}{4}}}
\Gamma\left(\frac{3}{4},ar_{\rm L}^4\right)
\right]
\end{equation}
and
\begin{equation}\label{eq:28e90jdid}
P_{\rm ave}^{\rm L}
\simeq 
P_0
\frac{\rho_{0}^{'}}{4a^{\frac{1}{4}}}
e^{br_{\rm L}^2/2}
\left[
\frac{1}{\sqrt{a}r_{L}^2}
\Gamma\left(\frac{3}{4},ar_{\rm L}^4\right)-\Gamma\left(\frac{1}{4},ar_{L}^4\right)
\right].
\end{equation}
Even though these results are also not completely transparent, it is evident from the equation above that $P_{\rm ave}^{\rm L}$ will be much smaller than $P_0$ for reasonable choices of the parameters because $\rho_{0}^{'}$ is in the order of $r_{\rm L}$, $a^{\frac{1}{4}}$ is much larger than $r_{\rm L}$, and the remaining terms are either order of 1 or smaller. To estimate $P_{\rm ave}^{\rm L}$ for $ar_{\rm L}^4\gg 1$ and $P_0 \kappa_{\rm P} r_{\rm L}^2\gg 4k_{\rm B}T$, we can use the asymptotic series\cite{abramowitz1964handbook}
\begin{equation}      
\Gamma\left(a,x\right) = x^{a-1}e^{-x}\left(1+\frac{a-1}{x} + ... \right),
\end{equation}
which is valid for large $x$. Retaining the first two terms of the series, we obtain
\begin{equation}
\Gamma\left(\frac{3}{4},ar_{\rm L}^4\right) \simeq e^{-ar_{\rm L}^4} \frac{1}{a^{1/4}r_{\rm L}}
\left(1-\frac{1}{4ar_{\rm L}^4}\right),
\end{equation}
\begin{equation}
\Gamma\left(\frac{1}{4},ar_{\rm L}^4\right) \simeq e^{-ar_{\rm L}^4} \frac{1}{a^{3/4}r_{\rm L}^3}
\left(1-\frac{3}{4ar_{\rm L}^4}\right).
\end{equation}
The substitution of the above equations into Eq.~(\ref{eq:28e90jdid}) gives the lower bound for $P_{\rm ave}$ in Eq.~(\ref{eq:PaveL}):
$$
P_{\rm ave}^{\rm L}
\simeq 
\frac{2/P_0 }{
1
+\frac{1}{4a r_{\rm L}^4}
\left(1-\frac{3}{4ar_{\rm L}^4}\right)
}
\left(
\frac{k_{B}T}{\kappa_{\rm P} r_{\rm L}^2}
\right)^2
\simeq 
\frac{2}{P_0 }
\left(
\frac{k_{B}T}{\kappa_{\rm P} r_{\rm L}^2}
\right)^2.    
$$

\subsubsection*{Exposure to laser power: upper limit (Eq.~(\ref{eq:PaveU})).}

To calculate an upper limit to the laser power, we assume a uniform distribution of the particle position. We make use of the fact that, while $\rho(z)$ extends to infinity, it drops extremely fast as a result of the quartic term of $r$ within the exponential. This creates an excellent opportunity for the calculation of an approximate result and suggests a well-defined point to terminate the uniform distribution. To this end, we introduce a new parameter, $r_{\rm max}$, which represents a finite amount of displacement beyond which the probability of finding the particle is taken to be negligible. More precisely, at $r_{\rm max}$, the probability is $1/\mathcal{C}$, where $\mathcal{C} \gg 1$. This value is given by $e^{-ar_{\rm max}^4+br_{\rm max}^2-br_{\rm L}^2/2}=\mathcal{C}^{-1}$, which is easily solved to yield
\begin{equation}
r_{\rm max} 
=
r_{\rm L}
\sqrt{
1+\sqrt\frac{4 \ln{(\mathcal{C})}k_{\rm B}T}{r_{\rm L}^2\kappa_{\rm P} P_0 }
}~. 
\end{equation}
While the choice of $\mathcal{C}$ is arbitrary, $r_{\rm max}$ is remarkably insensitive to this choice due to the presence of the logarithm, which is furthermore imbedded within two square roots. The utility of this approach derives from this extreme insensitivity. 

With these simplifications, the upper bound to the variance of the particle position can be calculated as
\begin{equation}
\sigma_{{\rm U},r}^2
=
\frac{1}{r_{\rm max}}\int_{0}^{r_{\rm max}}r^2 {\rm d}r
=
\frac{1}{3}r_{\rm max}^2
\end{equation}
and that to the average power to which the particle is exposed is given by
\begin{equation}
\begin{split}
P_{\rm ave}^{\rm U}
& =
\frac{1}{r_{\rm max}}\int_{r_{\rm L}}^{r_{\rm max}}P_0 
\left(
\frac{r^2}{r_{\rm L}^2} - 1
\right) {\rm d}r \\
& = 
P_0 
\left(
r_{\rm max}-r_{\rm L}
\right)
\left(\frac{r_{\rm max}}{3r_{\rm L}^2}+\frac{1}{3r_{\rm L}}-\frac{2}{3r_{\rm max}}\right).
\end{split}
\end{equation}

It is important to remember that the uniform distribution is not an exact upper limit to the exact result by virtue of the region from $r_{\rm max}$ to $\infty$. However, it is straightforward to show that this region has a negligibly small contribution due to the rapidly decaying  $\rho(r)$ for $r>r_{\rm max}$  for sufficiently large $r_{\rm max}$. Namely, we want to show that $\int_{r_{\rm L}}^{r_{\rm max}}e^{-ar^4+br^2}{\rm d}r 
\gg 
\int_{r_{\rm max}}^{{\rm \infty}}e^{-ar^4+br^2}{\rm d}r$. To this end, we introduce $S(r)$,
\begin{equation}
S(r)=\int_{r_{\rm max}}^{r}e^{-ar^4+b^2}{\rm d}r~.
\end{equation}
Using the definitions of $a$ and $b$, $S(r)=\int_{r_{\rm max}}^{{\rm r}}e^{-ar^2(r^2-2r_{\rm L}^2)}{\rm d}r$. For $r> \sqrt{2} r_{\rm L}$, $ar^2(r^2-2r_{\rm L}^2)>ar^2$, and $S(r)<\int_{r_{\rm max}}^{r}e^{-ar^2}{\rm d}r=\frac{\sqrt\pi}{\sqrt a}
\left(
{\rm erf}(\sqrt ar)- {\rm erf} (\sqrt ar_{\rm max})
\right)$. Therefore, as a limiting case, we obtain 
\begin{equation}
\begin{split}
\lim_{r\to\infty}S(r)
& <\lim_{r\to\infty}\frac{\sqrt\pi}{\sqrt a}
\left(
1-{\rm erf}(\ln{\mathcal{C}})
\right) \\
& \simeq \frac{\sqrt\pi}{\sqrt a}\frac{e^{-(\ln{\mathcal{C}})^2}}{\mathcal{C}}\simeq 0\; {\rm for}\; \mathcal{C} \gg 1~.
\end{split}
\end{equation}
It is clear that the contribution to power exposure from the region of $r>r_{\rm max}$, which will be neglected in the calculation of the approximate upper limit is, bounded, small in absolute terms, and significantly smaller than the overestimation that arises from the region $r_{\rm L} < r < r_{\rm max}$ for any experimentally relevant choice of the parameters. 

In the case of $r_{\rm L} \lesssim r_{\rm max}$, the average power exposure is given by
\begin{equation}
\begin{split}
P_{\rm ave}
& =\frac{1}{r_{\rm max}}\int_{r_{\rm L}}^{r_{\rm max}}P_0 
\left(
\frac{r^2}{r_{\rm L}^2}-1
\right){\rm d}r \\
& =
\frac{P_0 }{3r_{\rm L}^2r_{\rm max}}\left(r_{\rm max}^3-r_{\rm L}^3\right)-\frac{P_0 }{r_{\rm max}}\left(r_{\rm max}-r_{\rm L}\right).
\end{split}
\end{equation}
Introducing $\Delta \equiv r_{\rm max}-r_{\rm L} \ll r_{\rm L}$, and using the Taylor expansion, 
\begin{equation}
 r_{\rm max}^3=r_{\rm L}^3
\left(1+\frac{\Delta}{r_{\rm L}}
\right)^3 \simeq
r_{\rm L}^3
\left(1+3\frac{\Delta}{r_{\rm L}}+3\left(\frac{\Delta}{r_{\rm L}}\right)^2+...
\right),
\end{equation}
we obtain the upper bound to the average power given by Eq.~(\ref{eq:PaveU}), i.e.,
$$
P_{\rm ave}^{\rm U}
=
P(r_{\rm max})\frac{\Delta}{2r_{\rm max}}
=
P(r_{\rm max}) \sqrt
{\ln{(\mathcal{C})}
\frac{ k_{\rm B}T}{4r_{\rm L}^2\kappa_{\rm P} P_0 }}~.
$$

\subsection*{Rate equations for the laser dynamics} 

The fiber laser in our experiment consist of an Yb-doped fiber section of length $L_{\rm g}$. To calculate the laser power circulation in the ring cavity of the laser, we solve the rate equations \cite{kelson1998strongly}. The governing equations for the pump power, $P_{\rm p}$, signal power $P_{\rm s}$, and amplified spontaneous emission, $P_{\rm ASE}$, in a given longitudinal position $Z$ at the $m^{\rm th}$ round trip are:
\begin{equation}
\frac{dP_{{\rm p},m}(Z)}{dZ}
=
\Gamma_{\rm p} 
\left( \sigma_{\rm p}^{\rm e} N_{2,m}(Z) 
- \sigma_{\rm p}^{\rm a} N_{1,m}(Z) \right)
 P_{{\rm p},m}(Z)
- \alpha _{\rm p} P_{{\rm p},m}(Z),
\end{equation}
\begin{equation}
\frac{dP_{{\rm s},m}(Z)}{dZ}
=
\Gamma_{\rm s} 
\left( \sigma_{\rm s}^{\rm e} N_{2,m}(Z)
- \sigma_{\rm s}^{\rm a} N_{1,m}(Z) \right)
P_{{\rm s},m}(Z)
- \alpha _{\rm s} P_{{\rm s},m}(Z),
\end{equation}
\begin{equation}
\begin{split}
\frac{dP_{{\rm ASE},m}^{\pm}(Z)}{dZ}
=
& 
\pm\Gamma_{\rm s} 
\left( \sigma_{s}^{{\rm e},\pm} N_{2,m}(Z)
- \sigma_{\rm s}^{{\rm a},\pm} N_{1,m}(Z) \right)
 P_{{\rm ASE},m}^{\pm}(Z) \\
& \pm 2 \sigma _{\rm s}^{{\rm e},\pm} \Gamma_{\rm s} \frac{h c^2 }{\lambda_{\rm s}^3}\Delta \lambda_{\rm s}  N_{2,m}(Z),
\end{split}
\end{equation}
where $\sigma^{\rm a(e)}_{\rm s}$,  $\sigma^{\rm a(e)}_{\rm p}$ are the absorption (emission) cross sections at signal and pump wavelengths, respectively; 
$\Gamma_{\rm p}$ and {$\Gamma_{\rm s}$ stand for pump and signal filling factors; 
$h$ is the Plank's constant; 
$\lambda_{\rm s}$ and $\Delta \lambda_{\rm s}$ are the signal wavelength and spectral width;
$N_{\rm 2}$ and $N_{\rm 1}$ are the population densities of the upper and lower lasing levels at $Z$ and are given by
\begin{widetext}
\begin{equation}
\frac{N_{2,m}}{N_{1,m}}
=
\frac{
R_{{\rm a},m} + W_{{\rm a},m} + W_{{\rm a},m} + A_{{\rm ASE},m}^{+} + A_{{\rm ASE},m}^{-}
}{
R_{\rm a} + R_{{\rm e},m} + W_{{\rm a},m} + W_{{\rm e},m} + W_{{\rm a},m} + W_{{\rm e},m} + A_{{\rm ASE},m}^{+} + A_{{\rm ASE},m}^{-} + A_{\rm s}
},
\end{equation}
\end{widetext}
where
$A_{\rm s}$ is the spontaneous transition rate,
$R_{{\rm a(e)},m} = P_{{\rm p},m} \frac{\Gamma _{\rm p}\sigma ^{\rm a(e)}_{\rm p}} {\nu_{\rm p}}$ are the pump transition probabilities, 
$W_{{\rm a(e)},m} = P_{{\rm s},m} \frac{\Gamma _{\rm s}\sigma ^{\rm a(e)}_{\rm s}}{\nu_{\rm s}}$ are the signal transition probabilities, 
and  $A_{{\rm ASE},m}^{\pm} = \sum\limits_{\lambda} P_{{\rm ASE},m}^{\pm} \frac{\Gamma _{\rm s}\sigma ^{{\rm e},\pm}(\lambda)}{\lambda} $.
In the ring cavity, a portion of the output power is fed back to the cavity for the next round trip and satisfy the following boundary condition:
\begin{equation}
P_{{\rm s},m+1}(0)= (1-l_{\rm scat})(1-l_{\rm cavity})P_{{\rm s}, m}.(L_{\rm g}),
\end{equation}
where $l_{\rm cavity}$ and $l_{\rm scat}$ represent optical loss of the cavity and the particle, respectively.
The power in the cavity effectively reaches to a steady state, $P$, after many roundtrips, i.e., for $m \gg 1$.

\subsection*{Numerical simulations}

To perform the numerical simulations, we project Eq.~(\ref{eq:ot}) on the Cartesian axes $x$, $y$ and $z$, to obtain a set of uncoupled overdamped Langevin equations, corresponding to the following system of coupled finite difference equations,\cite{volpe2013simulation}
\begin{equation}
x_i 
=
x_{i-1}
+ \gamma^{-1} F_{{\rm ot},x}{(x_{i-1},P_{i})} \Delta t
+ \sqrt{2\Delta t D} {w_{x,i}}~,
\end{equation}
\begin{equation}
y_i
= 
y_{i-1}
+ \gamma^{-1} F_{{\rm ot},y}{(y_{i-1},P_{i})} \Delta t
+ \sqrt{2\Delta t D} {w_{y,i}}~,
\end{equation}
\begin{equation}
z_i
= z_{i-1}
+ \gamma^{-1} \left[ F_{{\rm ot},z}{(z_{i-1},P_{i})} - F_{\rm g} \right] \Delta t
+ \sqrt{2\Delta t D} {w_{z,i}}~,
\end{equation}
where 
${\bf r}_i = [x_i,y_i,z_i]$ is the particle position at timestep $i$,
$P_i$ is the stationary power at timestep $i$,
${\bf F}_{\rm ot}({\bf r},P) = [F_{{\rm ot},x}(x,P),F_{{\rm ot},y}(y,P),F_{{\rm ot},z}(z,P)]$ is the optical force, 
$F_{\rm g}$ is the gravity minus the buoyancy acting on the particle, 
and $w_{x,i}$, $w_{y,i}$ and $w_{z,i}$ are independent Gaussian random
numbers with zero mean and unitary variance. 
The incident power on the particle, $P_i$, is updated at each timestep. 
The optical loss due to the trapped particle is obtained as $l_{{\rm scat},i}=1-P_{{\rm scat},i} / P_i$, where $P_{{\rm scat},i}$ is the power of the scattered light collected by the second collimator. 
This information is used together with the laser power rate equations to calculate the laser power at timestep $i$.

\subsection*{3D digital video microscopy} 

We have acquired videos of the particle at $50\,{\rm Hz}$ and then used standard digital video microscopy algorithms to detect its position \cite{crocker1996methods}. 
For the measurement of the lateral position ($r$) of the particle, we have used the standard centroid algorithm (lateral resolution $20\,{\rm nm}$).
For the measurement of the axial position ($z$) of the particle, we have acquired a stack of reference images of a stuck particle as a function of its $z$-position and compare them to the image of the particle in each frame (axial resolution of $40\,{\rm nm}$). 


\textbf{Acknowledgements.}
We thank Agnese Callegari and S. Masoumeh Mousavi for useful discussions.

\textbf{Funding}
This partially financed through the European Research Council (ERC) Consolidator Grant ERC-617521 Nonlinear Laser Lithography (NLL) awarded to F\"OI and the Starting Grant ERC-677511 ComplexSwimmers awarded to GV.

\end{document}